\documentclass[conference]{IEEEtran}
\IEEEoverridecommandlockouts
\usepackage{hyperref}
\usepackage{cite}
\usepackage{amsmath,amssymb,amsfonts}
\usepackage{algorithmic}
\usepackage{graphicx}
\usepackage{textcomp}
\usepackage{xcolor}
\usepackage{enumerate}
\usepackage{todonotes}

\usepackage{graphicx}
\usepackage{subcaption}
\usepackage{adjustbox}
\usepackage{mwe}
\usepackage{balance}

\def\BibTeX{{\rm B\kern-.05em{\sc i\kern-.025em b}\kern-.08em
    T\kern-.1667em\lower.7ex\hbox{E}\kern-.125emX}}
\begin{document}

\title{Robot Capability and Intention in Trust-based Decisions across Tasks
}

\author{
\IEEEauthorblockN{Yaqi Xie\IEEEauthorrefmark{1}, Indu P Bodala\IEEEauthorrefmark{1}, Desmond C. Ong\IEEEauthorrefmark{2}, David Hsu\IEEEauthorrefmark{1}, Harold Soh\IEEEauthorrefmark{1}}
\IEEEauthorblockA{\IEEEauthorrefmark{1}\textit{Department of Computer Science} \\
\textit{National University of Singapore}\\
Singapore \\
\{yaqixie, indu, dyhsu, harold\}@comp.nus.edu.sg
}
\\
\IEEEauthorblockA{\IEEEauthorrefmark{2}\textit{A*STAR Artificial Intelligence Initiative and Institute of High Performance Computing} \\
Singapore \\
desmond.c.ong@gmail.com
}
% \and
% \IEEEauthorblockN{Indu P Bodala}
% \IEEEauthorblockA{\textit{Computer Science} \\
% \textit{National University of Singapore}\\
% Singapore, Singapore \\
% indu@comp.nus.edu.sg}
% \and
% \IEEEauthorblockN{Desmond Ong}
% \IEEEauthorblockA{\textit{Institute of High\n Performance Computing} \\
% %\textit{Institute of High Performance Computing}\\
% Singapore, Singapore \\
% desmond.c.ong@gmail.com}
% \and
% \IEEEauthorblockN{David Hsu}
% \IEEEauthorblockA{\textit{Computer Science} \\
% \textit{National University of Singapore}\\
% Singapore, Singapore \\
% dyhsu@comp.nus.edu.sg}
% \and
% \IEEEauthorblockN{Harold Soh}
% \IEEEauthorblockA{\textit{Computer Science} \\
% \textit{National University of Singapore}\\
% Singapore, Singapore \\
% harold@comp.nus.edu.sg}
}

\maketitle

\begin{abstract}
In this paper, we present results from a human-subject study designed to explore two facets of human mental models of robots---inferred capability and intention---and their relationship to overall trust and eventual decisions. In particular, we examine delegation situations characterized by uncertainty, and explore how inferred capability and intention are applied across different tasks. We develop an online survey where human participants decide whether to delegate control to  a simulated UAV agent. Our study shows that human estimations of robot capability and intent correlate strongly with overall self-reported trust. However, overall trust is not independently sufficient to determine whether a human will decide to trust (delegate) a given task to a robot. Instead, our study reveals that estimations of robot intention, capability, and overall trust are integrated when deciding to delegate. From a broader perspective, these results suggest that calibrating overall trust alone is insufficient; to make correct decisions, humans need (and use) multi-faceted mental models when collaborating with robots across multiple contexts.
\end{abstract}


\section{Introduction}
Trust is a cornerstone of long-lasting collaboration in human teams, and is crucial for human-robot cooperation~\cite{Groom2007}. For example, human trust in robots influences usage~\cite{sheridan1984research}, and willingness to accept information or suggestions~\cite{freedy2007measurement}. Misplaced trust in robots can lead to poor task-allocation and unsatisfactory outcomes. In recognition of its importance, there has been a concerted effort in the research community to better understand the formation and dynamics of trust in robots and automation~\cite{meta,Muir1994,Muir1996,optimo,desai2012modeling,SohRSS18,chen}. 

Nevertheless, there remains crucial gaps in our understanding of human-robot trust, particularly in the role of inferred robot ``\emph{intention}'', i.e., what people believe  the robot is trying to achieve. Prior research has focussed largely on inferred \emph{capability}~\cite{optimo,chen,SohRSS18}, which has been shown to be a primary antecedent to trust~\cite{Hancock2011,schaefer2016meta}. However, with the advancement of robot technology (e.g., artificial intelligence), robots are increasingly poised to achieve peer-like collaboration with humans. In this new role, robots may be afforded greater autonomy, which involves independent decision-making, and trust in intention may surface as a critical factor~\cite{Groom2007,Castelfranchi2007,Huangb}. 

 %However, the same passenger may value his life over the lives of pedestrians, but the AV may value all human life equally.    
%However, in any given context, a robot and human may \emph{not} be value-aligned. For example, a passenger in an autonomous vehicle (AV) 

In this paper, we seek to clarify the role of \emph{both} inferred robot intention and capability in trust-based scenarios. Inspired by prior work on inter-human and socio-cognitive  trust~\cite{mayer1995integrative,Castelfranchi2007}, we posit that when deciding to delegate tasks to a robot, a user considers two complementary aspects: (i) whether the robot has the proper intention or motive, i.e., if it is optimizing the correct objectives, and (ii) whether the robot has sufficient capability, i.e., if it is able to carry out the task successfully under those objectives. For example, a passenger in an autonomous vehicle (AV) may trust in the AV's capability to navigate in a complex environment, yet distrust the AV's intention to hasten his arrival by circumventing road safety, or to value his life over the lives of others in an emergency.

We present results from a human-subject study ($n=400$) where participants had to choose whether to delegate control to a robot under varying conditions. In particular, we sought to examine how estimations of robot intention and capability developed under one task affected trust-based decisions in the same task, and \emph{subsequent novel tasks} (where arguably, trust plays a more important role). We created an online survey where participants had to decide whether to trust an Unmanned Aerial Vehicle (UAV) in three different tasks: (i) searching, (ii) mapping, and (iii) fire-fighting. After interacting with the robot in the searching task, participants had to decide whether to trust the robot in the latter two tasks. The searching and mapping tasks required similar robot capabilities, but involved potentially different objectives. Likewise, similar intentions (i.e., risk-behavior) would arise in the searching and fire-fighting tasks, but different robot capabilities are required. 

Our primary finding is that decisions to delegate control in novel task contexts depend not only on overall trust in the robot, but also on estimations of robot capability and intention. In other words, humans appear to integrate several facets of their mental model (i.e., their estimations and beliefs) to arrive at a decision to trust. Furthermore, our results suggest that inferred capability and intention transfer (or generalize) separately to new situations, which extends previous results~\cite{SohRSS18} and suggests avenues for future work in assessing the degree of transfer. 

These results suggest that human-robot trust may be similar to trust in humans, in that both inferred intention (motive) and capability play a role. However, trust in \emph{non-robot} automation (e.g., automated alarms and decision aids) has also been shown to develop similarly to trust in humans, but with critical differences~\cite{Madhavan2007}. For example, people perceive automated systems to be more credible and objective sources of information compared to humans (to the point where erroneous decisions are agreed with). However, people are less tolerant of automation errors, leading to sharp declines in trust when a mistake is perceived. Here, we compared trust-based decisions when participants were informed they were working with another human player, and when they were paired with a robot. In both these cases, the agent was a confederate software program with the \emph{same} behavioral policy. Interestingly, we did not find that humans always engaged in more trusting behavior when they thought they were working with another human (rather than a robot), \emph{despite} reporting significantly higher trust in those situations. This suggests a complex relationship between trust, observed performance, and prior expectations given the perceived agent type. 

To summarize, this work contributes a novel investigation into trust-based decisions in human-robot teams where robot capability and intention play a role in outcomes. We find evidence that humans utilize rich mental models of robot teammates when choosing to trust. Self-reported trust in the robot is an insufficient predictor of human decisions to trust the robot in different contexts. Rather, our results show that decisions to trust are based on human's mental model of the robot, which includes inferred capability, intention, and potentially other components. Furthermore, people's mental models appear to differ when working with a human partner or a robot---this points to a potential difference in prior expectations when working with humans versus machines. Taking a broader perspective, our results have implications for the design of human-centric robots that are able to reason about human trust and act accordingly~\cite{chen}. For example, robots that aim to ``teach'' humans to make appropriate delegation decisions should calibrate not only general trust, but also their estimations of the robot's intention and capabilities.

\section{Preliminaries: Background and Related Work}

To situate our work, we briefly review trust research, which is a large interdisciplinary endeavor spanning multiple fields including human-factors, social science, and human-robot interaction. Trust has been studied in many forms: trust in other humans, trust in organizations, and trust in machines. Here, we discuss the literature with a focus on \emph{human trust in robots} and the aspects most relevant to this paper, i.e., the key definitions and concepts, and research in modeling trust in robots.

\subsection{Characterizing Trust: Concepts and Definitions} 
Trust is a concept with varying definitions even within the same field. For example, trust has been defined as a belief---the subjective probability whereby an agent (the \emph{trustor}) assesses whether another agent (the \emph{trustee}) will perform an action~\cite{gambetta2000can}. However, this definition has been criticized for lacking task context: for example, it fails to take into account the risk associated with the task at hand~\cite{Castelfranchi2007}. An alternative risk-related definition of trust is as the belief that a trustee will help the trustor's goal in a situation characterized by uncertainty and trustor vulnerability~\cite{reliance}, or just the willingness of the trustor to depend on another agent, even with the risk of possible negative consequences~\cite{josang2007survey}. In this work, we adopt a recent definition of trust used in human-robot interaction, i.e., that trust is a latent (hidden) variable that summarizes past experience with an agent/robot~\cite{chen,SohRSS18}, which is useful for predicting future behavior of the trustee and making a decision to put oneself in a position of vulnerability. By summarization, we mean a mental abstraction or model of past interactions that is predictive of the trustee's future behavior, since  humans generally do not remember the entirety of past interactions with other agents. %, but maintain a mental model that contributes to behaviors when interacting with the agent in novel situations. 

There are two types of trust that differ in their situation specificity: on the one hand, there is \emph{dispositional trust} or \emph{trust propensity}, which is an individual difference for how willing one is to trust another. Some people may inherently be more trusting~\cite{Yamagishi1999}. On the other hand, \emph{situational} or \emph{learned trust} results from interaction between the agents concerned. The more you use your new robot, the more you may learn to trust it. Dispositional trust is a trustor \emph{trait}, i.e., it differs between trustors, and tends to be similar for the same trustor across different situations. By contrast, situational trust is a trustor \emph{state}, i.e., it is specific to the task at hand, and may change frequently given new information. In this work, we will be concerned primarily with situational trust in robots, but we bear in mind individual differences in dispositional trust when examining empirical data.

%An operationalized definition of trust is as ``a belief, held by the trustor, that the trustee will act in a manner that mitigates the trustor’s risk in a situation in which the trustor has put its outcomes at risk''~\cite{Wagner2009}. More recently, trust has been viewed as latent variable that summarizes past experience with an agent/robot~\cite{chen,SohRSS18}.

%If we examine the literature further, one facet of trust concerns the notion of trustee ability, or the competence of the trustee to perform its task. Do you trust your doctor to perform your surgery? Another attribute of trust is benevolence, i.e., the extent to which the trustee acts to benefit the trustor. Do you trust your lawyer to act in your best interests? And finally, trust in agent integrity, which is the belief that the trustee adheres to moral and ethical principles, e.g., values relating to fairness. Do you trust your politicians to be honest? We group the latter two, benevolence and integrity, under “trustworthiness” or intent, as they relate to the motivation of the trustee in helping the trustor.

\subsection{Trust in Automation and Robots}
%Trust in automation (e.g., airplanes, plant process control) differs from trust between people in that automation is thought to lack inherent \emph{intentionality}~\cite{reliance,lewandowsky2000dynamics}. 
Muir's seminal work on trust in automation~\cite{Muir1994,Muir1996} and its effect on operator control allocation found that people's reported subjective trust ratings were sensitive to the automation's properties---the more reliable the automation, the higher the trust ratings, which subsequently led to increased automation reliance. In other words, human trust was influenced by machine behavior, which in turn influenced human behavior. Lee and Moray~\cite{LEE1992} found that this relationship was moderated by the operator's self-confidence, i.e., if trust in automation was greater than the operator's self-confidence, the operators were more likely to rely on the machine. 

As AI technology has matured, the research community has examined the trust in intelligent systems and robots, such as autonomous vehicles~\cite{SohRSS18,Huanga}, unmanned aerial vehicles (UAVs)~\cite{optimo} and medical diagnosis systems~\cite{Carlson2011}. In general, trust in human-robot interactions can be influenced by robot-related factors (e.g., performance, physical attributes), human-related factors (e.g., workload, self-confidence), and environmental factors (e.g., group composition, culture, task type)~\cite{Hancock2011}.  A majority of prior work has focussed on performance-related factors, particularly robot capabilities; for example, Soh \emph{et al.}~\cite{SohRSS18} examined the dynamics and transfer of trust in robot capabilities across tasks, where transfer is the ability to employ knowledge acquired in one task to improve performance in another~\cite{ShapiroKO08}. Recent work has explored the role of the robot's intention, e.g., its policy~\cite{Huangb,huang2018enabling} and decision-making process~\cite{Wang2016a}. This work adds to this body of literature and considers both intent and capability across tasks. 

%and trust can be utilized in decision making scenarios 

%Although a given robot may lack an inherent value system, people may anthropomorphize intelligent machines, and perceive intent or benevolence~\cite{Hancock2011,lee2010trust}. Yet, one can also argue that robot intentionality is constrained by the programmer, and one can make assumptions different from when dealing with other people (e.g., the robot was programmed with other humans’ best interests at heart). A crucial question is then to what degree perceived intentionality affects trust and whether this affects eventual decision to trust. 

% capability and intent definition

% \textbf{\color{red} the following is going to be really confusing for readers. I suggest changing the first two to ``true'' intention/capability. And the latter to ``inferred'' intention/capability. and being consistent throughout the paper}
%In our work, ``robot capability'' refers to the robot's ability to perform the task successfully, and ``robot intention'' refers to robot's motives, e.g. decision-making criteria and utility function.  

In the following, it is important to distinguish between the robot's \emph{true} capability/intent versus the human's \emph{estimation} of the robot's capability/intent. These two may differ, particularly when the human has had little experience with the robot. We will use the term \emph{inferred capability} to refer to the human's estimation of the robot's ability to perform the task successfully. Likewise, \emph{inferred intention} refers to the human's estimation of the robot's underlying motives or decision-making criteria, rather than the agent's true utility function. 

%In this work, we explore how human trust develops with varying robot ``intention'' and capability. 

%where the human and robot are value-aligned, i.e., both share the same objective metric of task performance (e.g.,~\cite{chen,optimo,desai2012a,Pierson2016,Pippin2014,Wang2016}). However, in any given context, a robot and human may \emph{not} be value-aligned. For example, a passenger in an autonomous vehicle (AV) may value his life over the lives of others, but the AV may value all human life equally. 

\subsection{Trust Measurement}

Trust is both latent (i.e., unobservable) and dynamic, which presents challenges for measurement. Multiple measurement scales have been proposed to quantify the degree of trust in a robot, including binary measures \cite{binary}, continuous measures~\cite{desai2012a,Xu2016,LEE1992}, ordinal scales \cite{Muir1989,Hoffman2013,Jian2000} and an Area Under Trust Curve (AUTC) measure\cite{desai2012a} that captures participant's trust through the entire interaction with the robot by integrating binary trust measures over time. In this work, we use both self-reported trust measures (e.g., Schaefer's trust scale\cite{schaefer}) and behavioral measures (decisions to delegate tasks to the robot). 

\section{Human-Subject Study: Intention, Capability, and Trust}
The overarching goal of our experiments was to explore the relationship between the inferred intention, capability, overall trust, and the decision to trust. We designed a user study with a two \{\textbf{G}rouped with \textbf{R}obot vs. \textbf{G}rouped with \textbf{H}uman\} by four \{robot: \textbf{H}igh~\textbf{C}apability/\textbf{L}ow~\textbf{C}apability + robot: \textbf{H}igh~\textbf{R}isk-taking/\textbf{L}ow~\textbf{R}isk-taking intention\} between-subjects factorial design. 
%In the following, we describe our online platform and the robot agents used in our study. We also discuss our experimental methodology and the primary hypotheses under investigation. 

%In this section, we describe our human subjects study, which was designed to evaluate: 1)Human trust dynamic. 2) Relation between overall trust and capability, intent. 3) Human's decision making process. 4) The transfer of trust across scenarios and tasks. 5) Difference between Human-Human trust and Human-Robot trust.

\subsection{Experimental Design}

\begin{figure}
\centerline{\includegraphics[width=0.98\columnwidth]{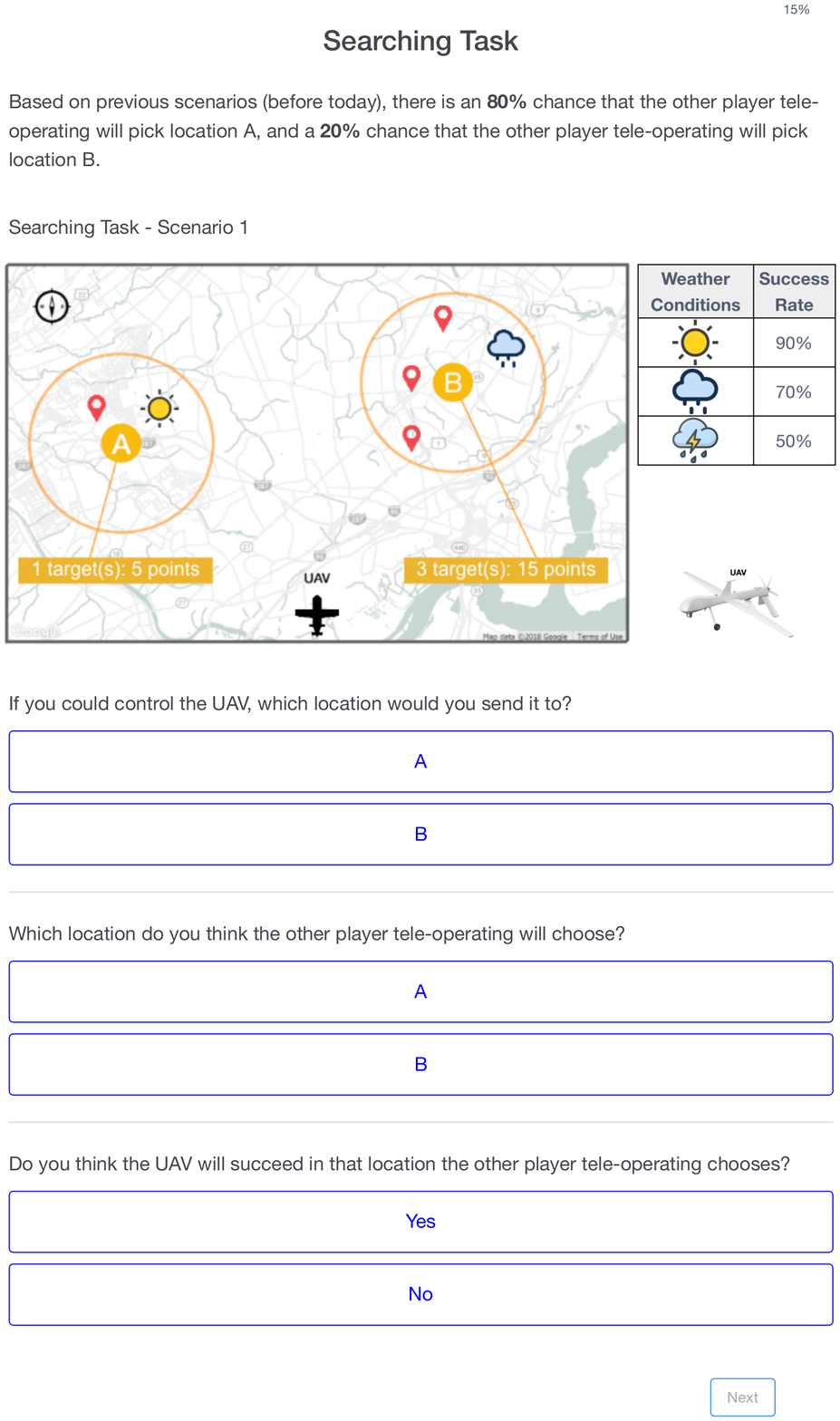}}
\caption{Example Scenario in Task 1 (Searching).}
\label{fig:simplatform}
\end{figure}

\noindent\textbf{Data Collection Platform and Tasks.} For this work, we developed an online survey (Fig. \ref{fig:simplatform}) with a delegation game; a human teammate decides whether to delegate control to an Unmanned Aerial Vehicle (UAV) performing three different mission tasks. Depending on their choice and the outcome of the task, they would gain a certain number of points. Choosing to take over control, i.e. tele-operate, would always cost some points, but this could be offset if the mission was a success. The three tasks were:
\begin{enumerate}
\item \textbf{Searching:} The UAV searches for targets by taking photos in various weather conditions. The UAV can choose between two locations, A and B, each with a different weather condition and number of targets. The weather condition affects the probability of success, which is shown to the participants (see Fig. \ref{fig:simplatform}).
\item \textbf{Mapping:} UAV creates maps by taking photos of the terrain. The UAV again chooses location A or B, but the weather in both locations are the \emph{same}. The difference is that visiting location A results in a low-accuracy but wide coverage map. Conversely, visiting B results in a high-accuracy, low coverage map. 
\item \textbf{Fire Fighting:} UAV puts out fires by dropping water/chemicals, again with two choices A and B.
The two locations would have different fire intensities and numbers of targets.
\end{enumerate}

In order to test the human trust transfer, we designed two types of capabilities and two types of intent:
\begin{itemize}
\item Capability 1 (Weather): Taking photos in various weather conditions. (Task 1 \& 2)
\item Capability 2 (Fire-fighting): Fire-fighting in various fire conditions. (Task 3)
\item Intent 1 (Risk preference): Lower risk but less reward vs. Higher risk but more reward. (Task 1 \& 3)
\item Intent 2 (Accuracy preference): Higher coverage but low accuracy vs. Lower coverage but higher accuracy. (Task 2)
\end{itemize}
Task 1 and Task 2 required the same robot capability but different decision-making criteria (intention), while Task 1 and Task 3 involved the same intention (risk taking behavior) but different capability.   

The tasks were designed to test how trust and learned mental models  transfer to situations where the robot capabilities required would differ (taking photos vs. fighting fires) or where potential differences in value assignments may occur (number of targets vs. mapping). With regard to the latter, participants were incentivized to pursue certain goals via point allocations, but were informed that the other agent did not necessarily optimize the same criteria. In each task, the robot either succeeded, whereby the participant would obtain all the stated points, or failed and the participant would receive  nothing (or negative points if they chose to teleoperate the robot). 

\vspace{0.5em}

\noindent\textbf{Confederate Agent/Robot Types.} In this study, human participants played with software agents, similar to \cite{Collins2016}. To investigate if trust differed significantly when participants thought they were playing with another person or a machine, we randomized the participants into two groups. In (\textbf{G}roup: \textbf{H}uman, \textbf{GH}), participants were informed that they were paired with another player who would tele-operate the UAV, i.e. make location decisions, should participants choose to delegate control. In (\textbf{G}roup: \textbf{R}obot, \textbf{GR}), participants were told they would interact with a robot. In both groups, participants interacted with one of four agent \emph{types} which differed along two dimensions: robot capability and intention. 

In Task 1, the robot's \emph{capability} was its ability to complete a mission in different weather conditions (clear, rainy, or thunderstorm). The success probabilities for the High Capability (HC) and Low Capability (LC) robots are shown in Table \ref{tbl:successrates}. The robot's \emph{intention} was related to its preference for risk; high risk-taking (HR) agents would attempt to maximize the number of targets even at locations where the chance of failure was high. Conversely, low risk (LR) agents are risk-averse. The exact decision made in the different scenarios were obtained via expected utility maximization, i.e., the agents would choose actions that maximized their expected utility:
\begin{align}
	\mathbb{E}[U] = \sum_o U(o)p(o|a) 
\end{align}
where $U$ is a utility function and $p(o|a)$ is the probability of an outcome $o$ given an action $a$. We used an exponential utility function:
\begin{align}
	U(o)=
\begin{cases}
\frac{(1-e^{-ao})}{\alpha} & \alpha \neq 0\\
\alpha & \alpha = 0
\end{cases}
\end{align}
which is commonly applied in economics to model risk propensity. The parameter $\alpha$ controls the degree of risk preference: $\alpha > 0$ for risk aversion, $\alpha = 0$ for risk neutral, and $\alpha < 0$ for risk seeking. In our work, for the risk-taking robot (HR), $\alpha_{\textrm{HR}} =-0.2$, and for the risk-averse robot (LR),  $\alpha_{\textrm{LR}} =0.8$. Note that outcomes were not shown for Tasks 2 and 3 so, neither capability specifications nor decision-making process were required.

\begin{table}
\begin{center}
\caption{Success Rates for Task 1 (Searching) in Different Weather Conditions for Agents with High or Low Capability.}
\label{tbl:successrates}
\begin{adjustbox}{width=0.28\textwidth}
\begin{tabular}{l | c c}
\hline
 & \multicolumn{2}{c}{\textbf{Capability}}\\
\textbf{Weather Type} & \textbf{High} & \textbf{Low} \\
\hline
\hline
\textbf{Clear}& 100\% & 66\%\\
\textbf{Raining}& 83\% & 0\%\\
\textbf{Thunderstorm}& 66\% & 0\% \\
\hline
\end{tabular}
\end{adjustbox}
\label{tab1}
\end{center}
\end{table}

\begin{figure}
\begin{center}
\includegraphics[width=0.85\columnwidth]{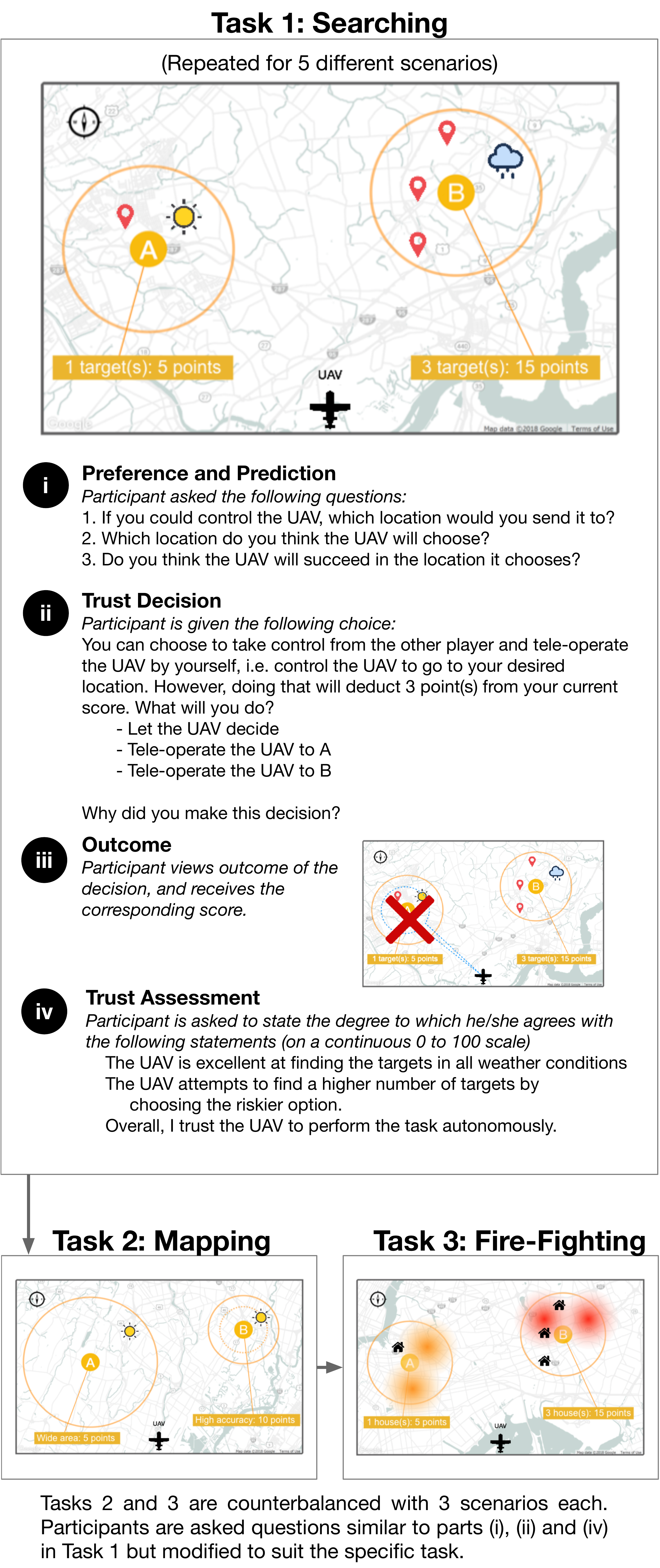}
\end{center}
\caption{Experiment workflow comprising three tasks.}
\label{fig:expworkflow}
\end{figure}

\subsection{Methodology}

\noindent\textbf{Participant Recruitment.} We recruited 400 participants  via Amazon Mechanical Turk (AMT). Participants were required to have at least a 99\% acceptance rate and were only allowed to participate in our survey once. Each survey lasted 30 minutes and participants were compensated \$3. Participants that attained a higher number of points had an opportunity to be compensated an extra \$10, which incentivized participants to pay attention and carefully consider their judgments.
\vspace{0.5em}
% In order to gather a large variety of participants, crowdsourcing (via Amazon's Mechanical Turk service) was used to collect data the experiment. In order to ensure the best possible data, participants were required to have at least a 99\% acceptance rate for their previous work and were only allowed to participate once. \par
%We have 4 robot types and 2 versions (human-robot, human-human), so there will be 8 groups in total. We recruited 50 people for each group, and therefore 400 participants in total. \par

\noindent\textbf{Procedure.} After providing consent and standard demographic data, participants were randomized to the two groups (GH or GR) and one of the four agent types. They were then presented with a description of Task 1 (Searching), and required to answer attention check questions. Participants were required to answer all questions correctly to proceed with the experiment. They were then allowed to play a trial scenario until they were ready to proceed. 

The remainder of experiment workflow is shown in Fig \ref{fig:expworkflow}. The participants were presented with five different scenarios for Task 1; each scenario consisted of a different number of targets (and hence, achievable points) and weather conditions in locations A and B. Each scenario description was followed by a sequence of four stages:
\begin{enumerate}[(i)]
	\item{Preference and Prediction:} Participants were asked to indicate their preferred location, and to predict where the robot would choose to go and the chance of success.
	\item{Trust Decision:} They then had to choose whether to delegate control to the robot, or to perform a take-over by tele-operating the UAV and overriding the UAV's choice in GR or the other player's choice in GH.
	\item{Outcome:} The outcome of their choice would then been shown, along with the points they would receive.
	\item{Trust Assessment:} Participants would then be asked to provide agreement scores  on statements regarding the UAV's competence, risk-behavior, and their overall trust.
\end{enumerate}

Participants then proceeded to Task 2 (Mapping) and Task 3 (Fire-fighting). The order of Task 2 and Task 3 were counterbalanced to eliminate order effects. Each task comprised 3 scenarios. Similar to Task 1, each scenario was followed by (i), (ii) and (iv); we excluded stage (iii) since providing additional observations may change participants' mental models. After completing all three tasks, participants completed a short questionnaire regarding their trust in the robot, the inferred robot capability and risk-behavior.

%And the study was approved by Review Board (IRB).

\vspace{0.5em}

\noindent\textbf{Dependent Variables} The primary dependent variables consist of both subjective self-reported measures---e.g., of overall trust in the robot and estimations of its capability, intention---and an objective measure---whether they decide to allow the agent to perform the task autonomously. For the self-reported measures, participants indicated the agreement to statements via a continuous scale ranging from 0\% to 100\% where 0 indicated complete disagreement and 100 indicated complete agreement. Further details about the dependent measures are listed below:
\begin{itemize}
\item \textbf{Trust Decision.}  This objective behavioral measure  captures whether the participants trusted the agent to perform the task by itself. In our setup (see part (ii) in Fig. \ref{fig:expworkflow}), they chose either to let the UAV decide the intended location (a decision to trust), or to tele-operate the robot to a specific location (a decision \emph{not} to trust). 
\item \textbf{Self-reported Trust.} We asked patients to state their agreement with the statement, ``Overall, I trust the UAV to perform the task autonomously''. We also included eight questions from Schaefer's trust scale~\cite{schaefer}.
\item \textbf{Self-reported Inferred Capability and Intention Estimation.}  We measured participants' perception of the robot's capability in the three different tasks using agreement statements, e.g., ``The UAV is excellent at finding the targets in all weather conditions.'' for Task 1. Similarly, for risk-behavior assessment, we used the statement, ``The UAV attempts to find a higher number of targets by choosing the riskier option''. At the end of the survey, we asked participants to choose between one of four options about the type of agent they had interacted with in the survey: (1) ``Highly capable but tends to take risk.'', (2) ``Highly capable but is conservative.'', (3) ``Not very capable but tends to take risk.'', or (4) ``Not very capable and is conservative.''. 
\item \textbf{Robot Decision and Outcome Prediction} We measured the participants' predictions about the robot's choice and the outcome given the choice, via two binary  answers to the questions: ``Which location to you think the UAV will choose?'' and ``Do you think the UAV will succeed in the location it chooses?'', respectively. We also computed intent alignment by comparing the predicted robot decision to the decision the participant would have made (via their answer to the question ``if you could control the UAV, which location would you send it to?''). 
\item \textbf{Self-reported Decision-making Similarity.} This captured the degree of perceived similarity to the agent in terms of the decision making process, measured via the agreement statement, ``The [UAV's $|$ other player's] decisions are similar to mine.''%, for the GR group. 
\end{itemize}

%
%\begin{itemize}
%\item \textbf{Trust Decision}  This is a binary number that measures whether the participant trust the robot (not tele-operate) or dis-trust the robot (tele-operate). 
%\item \textbf{Self-reported Trust Score} This measured the degree the participant agrees with the statement "Overall, I trust the UAV to perform the task autonomously.", using a continuous scale ranging from 0\% to 100\%.
%\item \textbf{Self-reported Capability Score} This is to measure the participants' inference about the robot's capability in this task, measured using a continuous scale ranging from 0\% to 100\%.
%\item \textbf{Self-reported Intent Score} This is to measure the participants' inference about the robot's intent about this task, measured using a continuous scale ranging from 0\% to 100\%.
%\item \textbf{Robot Performance Prediction} This measure the participants' prediction about both the robot's choice and the outcome given the choice, via 2 binary values. 
%\item \textbf{Self-reported Similarity Score} This measure the degree of similarity to the robot the participants perceived in term of the decision making process.  The value ranges from 0\% to 100\%.
%\end{itemize}

\subsection{Hypotheses}
%\textbf{\color{red} Do we want to move these to the beginning of Sec III?}
% motivation of designing these hypotheses

Our overarching hypothesis is that humans infer both robot capability and intent from observations, and use these estimations to make decisions whether to trust the robot in new, but related, tasks. We specifically hypothesized that:
\begin{itemize}
\item H1: Humans infer capability and intention from observations of robot performance.
\item H2: Inferred capability and intention transfer separately to different tasks.
\item H3: Both inferred capability and intention contribute to self-reported trust in the robot.
\item H4: Both inferred capability and intention influence human decisions to trust the robot.
\item H5: People are more trusting of the simulated human agents, rather than the simulated robot agents.
\end{itemize}

Hypotheses H1 and H2 capture our expectation that the different  robot types and performance would engender different mental models, and that specific facets of the mental models would carry over to the other tasks. H3 and H4 are our primary hypotheses relating inferred capability and intent to trust (both self-reported and behavioral). Finally, H5 encodes our expectation that perceived agent type also has an effect on trust; recent work has found that humans tend to trust other humans more compared to intelligent decision aids~\cite{Madhavan2007} and software agents in economic games~\cite{Collins2016}, even when the behavior of the confederate agent was identical. %However, less work has been performed evaluating if this finding extend to robots, which have physical capabilities. 

% \textbf{\color{red} In the following, we talk about ``trust in capability/intention''. I think we have to make a decision here. Do we mean trust  has different facets? or the mental models have different facets? Is the model: observations $\rightarrow$ mental models (cap/intent/others)$\rightarrow$trust$\rightarrow$decision OR observations$\rightarrow$trust(cap/intent/others)$\rightarrow$decision or potentially something else?}.
%We are also interested in whether the transfer of inferred capability, inferred intention and overall trust transfer to new, unseen tasks differ according to tasks properties.

\section{Results}

Data from 400 participants (Mean age: 38.85 years, 48\% female) were included in the following analysis. Overall, we found that human estimations of robot capability and intent are important factors in determining overall trust and trust-based decisions. We analyzed the two groups GR (playing with a simulated robot) and GH (playing with a simulated human player) separately; each group contained responses from 200 participants. The statistical analyses for H1-H4 were performed using GR data only since they are defined over human-robot trust, while H5 analysis compares GR with GH. %Statistical comparisons were made primarily using the Mann-Whitney U test for differences in distributions.

\begin{figure}
\centerline{\includegraphics[width=0.45\textwidth]{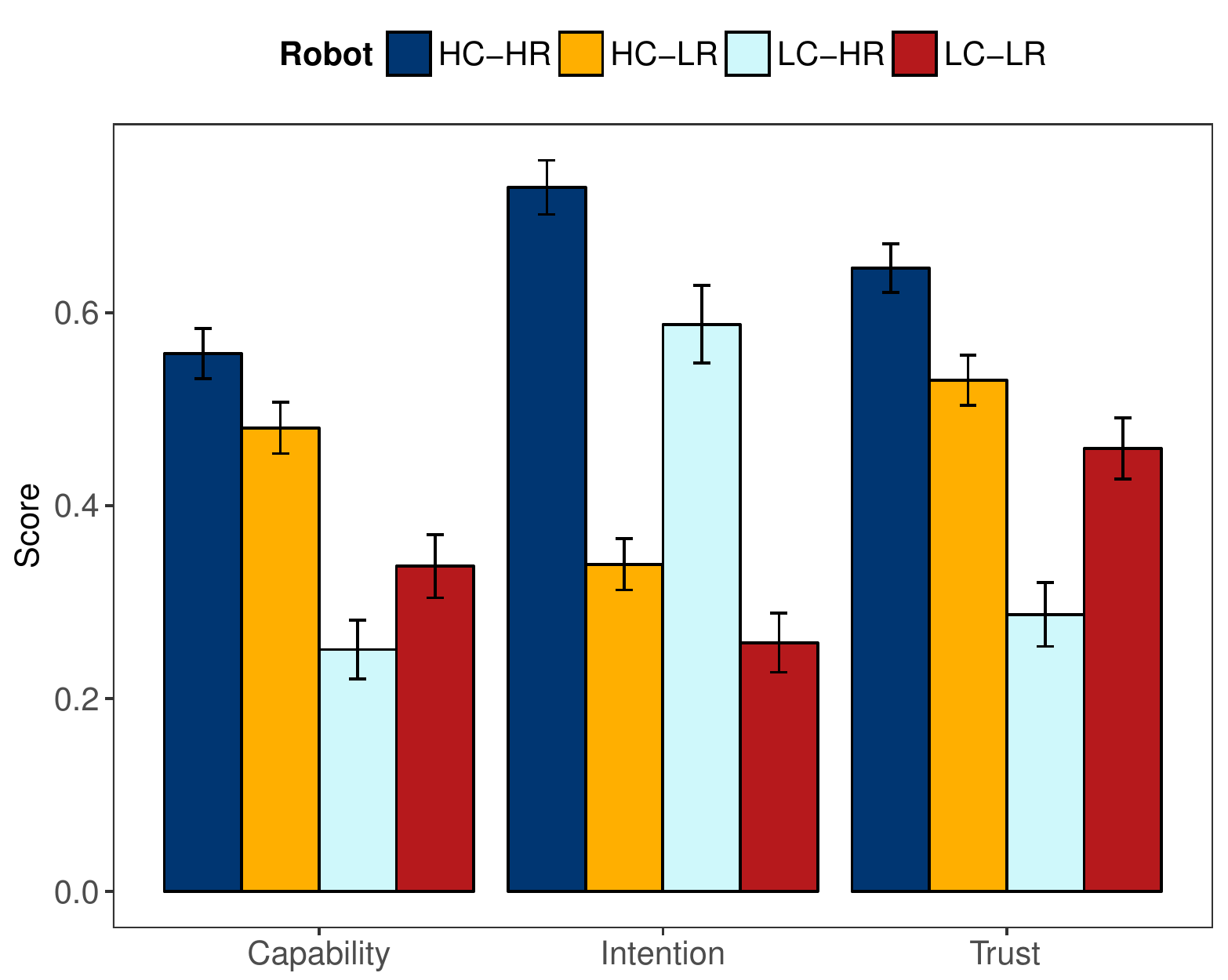}}
\caption{Self-reported Trust, Inferred Capability and (Risk) Intention scores for the four robot types: High-Capability and High Risk preference (HC-HR), High-Capability and Low Risk preference (HC-LR), Low-Capability and High Risk preference (LC-HR), Low-Capability and Low Risk preference (LC-LR). Participants reported higher capability scores for the HC robots, and higher risk intention for the HR robots, indicating they were able to infer  capability and risk preference from observed robot behavior and outcomes.}
\label{fig:inferred_trust_cap_intent}
\end{figure}

\vspace{0.5em}
\noindent\textbf{H1: Humans infer capability and intention from observations of robot performance.}
Fig. \ref{fig:inferred_trust_cap_intent} summarizes trust, inferred capability and intention (risk preference) scores for Task 1\footnote{Recall that the Searching task (Task 1) was used in a learning phase---outcomes are shown to participants, allowing them to update their mental representation of the robot.}. Participants reported different inferred capability and intention for each agent ($p<0.01$ across the four agent types). Specifically, inferred capability for HC-HR and HC-LR ($M = 0.52, \textrm{SE} = 0.02$) agents are significantly higher than LC-HR and LC-LR ($M = 0.29, \textrm{SE} = 0.02$); $t(398) = 7.62, p < 10^{-11}$. The inferred risk preference for HC-HR and LC-HR ($M = 0.66, \textrm{SE} = 0.03$) were significantly higher than HC-LR and LC-LR ($M = 0.42, \textrm{SE} = 0.03$); $t(398) = 5.99, p < 10^{-8}$. These results support H1. In addition, reported score differences between agents having the same true capability (or intent) were smaller compared to agents that differed in that dimension; the mean difference was $\approx 0.08$ for both intent and capability, compared to $\approx 0.2$ when the agent's true intent/capability differed. 

Participants were also largely able to discriminate between the robot types. As shown in Table \ref{tbl:robottypes}, there is a high agreement for all the cases except for LC-LR agent. Participants appeared to confuse the high capability and risk-averse agent, and the low-capability and risk-averse agent. This confusion is more severe in the simulated human group (GH). It was possible that since participants were allocated to only one agent type in our between-subjects design, they lacked a relative basis for comparison. Moreover, the low-capability and risk-averse robot would tend to choose the ``safer'' option, resulting in more successes.

%On the other hand, if we keep weather capability score only (or risk intent score only), we lose the ability to differentiate low risk-taking robots and high risk-taking robot (or high weather capability robot and low weather capability robot). Furthermore,  \par
%Note that HC-LR and LC-LR agents share similar trust scores, ($M_{HCLR} = 0.53, M_{LCLR}= 0.459; U = 987.5, p = 0.04$) but have distinct capability, ($M_{HCLR} = 0.48, M_{LCLR}=0.33; U = 737.5, p =0.0001 $), and intent scores, ($M_{HCLR} = 0.34, M_{HCLR} = 0.25; U= 903.5, p=0.008 $). Therefore, we lose the information to differentiate these two types if we keep trust score only. 
 
% \begin{figure}
% \centering
%   \begin{subfigure}{4.2cm}
%     \centering\includegraphics[width=4.2cm]{figure/h2-HC-HR}
%     \caption{HC-HR}
%   \end{subfigure}
%   \begin{subfigure}{4.2cm}
%     \centering\includegraphics[width=4.2cm]{figure/h2-HC-LR}
%     \caption{HC-LR}
%   \end{subfigure}
 
%  \centering
%   \begin{subfigure}{4.2cm}
%     \centering\includegraphics[width=4.2cm]{figure/h2-LC-HR}
%     \caption{LC-HR}
%   \end{subfigure}
%   \begin{subfigure}{4.2cm}
%     \centering\includegraphics[width=4.2cm]{figure/h2-LC-LR}
%     \caption{LC-HR}
%   \end{subfigure}
%   \caption{Self-reported overall trust, inferred capability and risk intent scores for 4 robot types for GR}
%   \label{fig:inferred_trust_cap_intent}
% \end{figure}

\begin{table}
\caption{Confusion Matrix of Predicted Robot Types.}
\label{tbl:robottypes}
\begin{center}
\begin{adjustbox}{width=0.45\textwidth}
\begin{tabular}{c c c c c}
\hline
\hline
\textbf{}&\multicolumn{4}{c}{\textbf{Simulated Robot Group (GR)}} \\
\textbf{Robot Type} & \textbf{\textit{HC-HR}}& \textbf{\textit{HC-LR}}& \textbf{\textit{LC-HR}} & \textbf{\textit{LC-LR}} \\
\hline
\textbf{HC-HR}& \textbf{0.65}& 0.2 & 0.15& 0\\
\textbf{HC-LR}& 0.25&  \textbf{0.575} & 0.075& 0.1\\
\textbf{LC-HR}& 0.125& 0.075 &  \textbf{0.7}& 0.1\\
\textbf{LC-LR}& 0.1& 0.225 & 0.15&  \textbf{0.525}\\
\hline
\hline
\textbf{}&\multicolumn{4}{c}{\textbf{Simulated Human Group (GH)}} \\ 
\textbf{Robot Type} & \textbf{\textit{HC-HR}}& \textbf{\textit{HC-LR}}& \textbf{\textit{LC-HR}} & \textbf{\textit{LC-LR}} \\
\hline
\textbf{HC-HR}&  \textbf{0.75}& 0.15 & 0.1& 0\\
\textbf{HC-LR}& 0.175&  \textbf{0.725} & 0.05& 0.05\\
\textbf{LC-HR}& 0.275& 0.025 &  \textbf{0.65}& 0.05\\
\textbf{LC-LR}& 0.1&  \textbf{0.525} & 0.125& 0.25\\
\hline
\hline
\end{tabular}
\end{adjustbox}
\label{tbl:confusion}
\end{center}
\end{table}

\vspace{0.5em}
\noindent\textbf{H2: Inferred capability and intention transfer separately to different tasks.} 
We expected inferred capability and intention to be separate components of a mental model that are transferred depending on the nature of the task. Recall that Task 2 (mapping) was designed such that the capability requirement of the robot was similar, i.e., operating in different weather conditions, but where value assignments were possibly different  compared to Task 1 (since the risk was the same at both locations). Hence, we expected estimations of capability to transfer (i.e., be similar), but not the estimations of intention (risk-preference). Conversely, Task 3 (fire-fighting) was designed such that the capability required was different, i.e., putting out fires rather than picture-taking in different weather conditions. However, choices in both Task 3 and Task 1 involved differences in risk. As such, we expected that estimations of intent to transfer, but estimations of capabilities not to transfer.

\begin{figure}
\centerline{\includegraphics[width=0.45\textwidth]{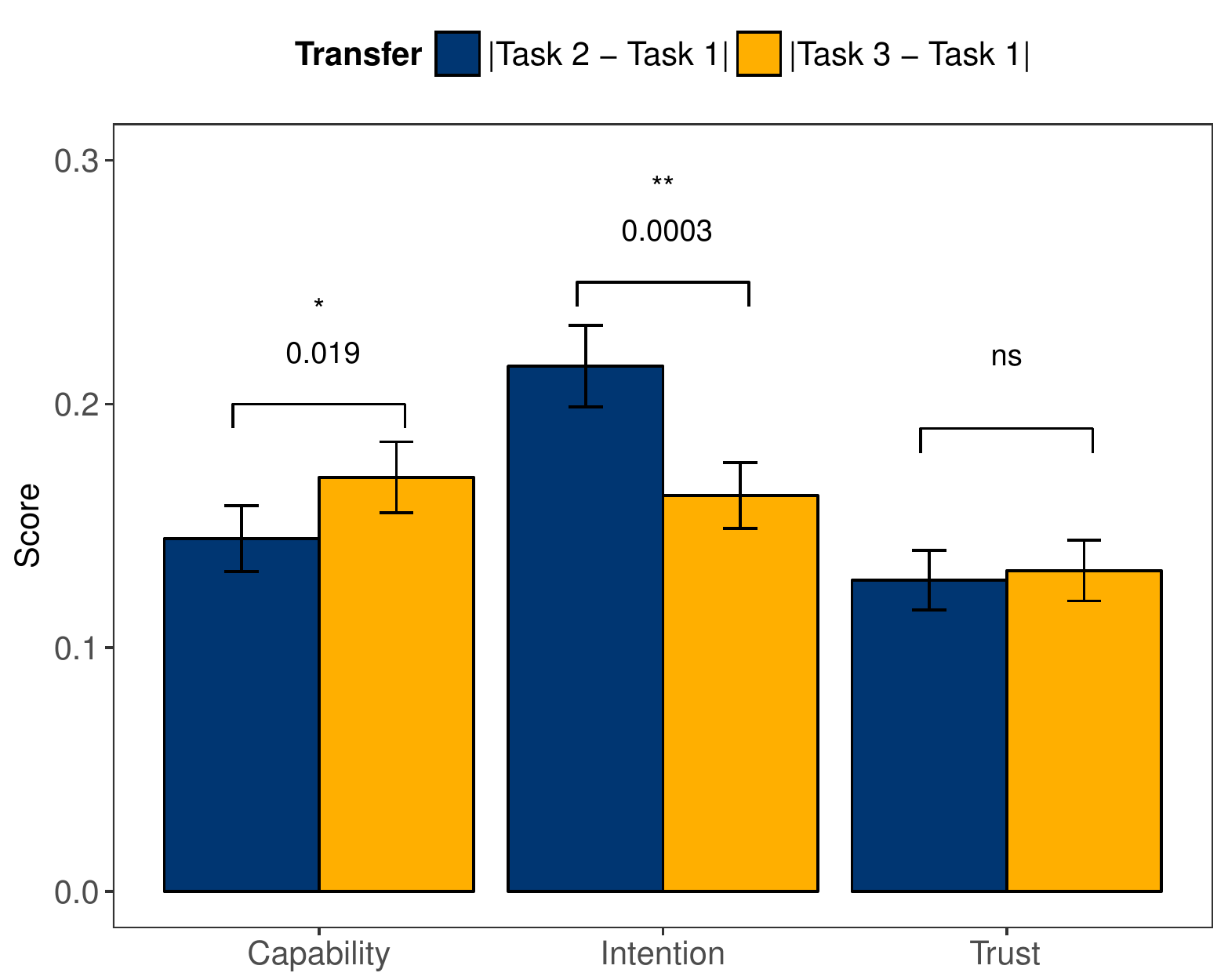}}
\caption{Transfer of self-reported trust, inferred capability and intent as measured by the absolute difference between scores ($p$-values shown above the bars). Differences in inferred capability and intent were significant between tasks, indicating differences in transfer depending on the task.}
\label{fig:transfer_all}
\end{figure}

We measured transfer across task contexts via the absolute difference of the  scores from Task 1 to Task 2 and from Task 1 to Task 3, similar to prior work~\cite{SohRSS18}. Lower differences indicate higher transfer. Fig. \ref{fig:transfer_all} summarizes our results for all robots, which support H2. As expected, inferred capability differences were significantly smaller (greater transfer) between Tasks 1 and 2 ($M = 0.14, \textrm{SE} = 0.01$), compared to Tasks 1 and 3 ($M = 0.17, \textrm{SE} = 0.01$); $t(199) = -2.37, p =0.02 $. Likewise, the transfer of inferred intent  was significantly greater between Task 1 to Task 2 ($M=0.22,\textrm{SE} = 0.01$), compared to Task 1 and Task 3 ($M= 0.16, \textrm{SE} = 0.01$); $t(199) = 3.72, p < 10^{-2}$. 

\vspace{0.5em}
\noindent\textbf{H3: Both inferred capability and intention contribute to self-reported trust in the robot.} We combined the responses from all scenarios in Tasks 1, 2 and 3, yielding a total of 2000 data points (200 participants $\times$ 10 scenarios) for analysis. Linear mixed models were used to account for repeated measures; inferred capability and intention were fixed main effects, with random intercepts for subjects, scenarios and tasks. Table \ref{tbl:linear_regression} summarizes our results; significant associations were found between inferred capability ($b = 0.42$, SE $=0.02$, $p <2\times10^{-16}$) and inferred intention ($b = 0.18$, SE $=0.02$, $p <2\times10^{-16}$) with self-reported trust scores, in support of the hypothesis.

%The results indicate a significant association between inferred capability ($b = 0.4086 (SE=0.02740)$, $p<10^{-10}$) and inferred intent ($b = 0.1539 (SE=0.02722)$, $p<10^{-7}$) with self-reported trust scores. The model results are summarized in Table \ref{tbl:linear_regression}.

\begin{table}
\begin{center}
\centering \caption {Mixed Effects Linear Regression Model relating Inferred Capability and Intention to Self-reported Trust.}
\begin{adjustbox}{width=0.45\textwidth}
\begin{tabular}{l c c c c}
\hline
\hline
% \multicolumn{3}{c}{\textbf{Trust, Capability, Intent to Trust Decision}}\\
\textbf{} & \textbf{Coef.} & \textbf{SE} & \textbf{t value} & \textbf{Pr$(>|t|)$}  \\
\hline
\textbf{Capability}&0.42  &0.02&23.39&$<2\times 10^{-16}$\\
\textbf{Intention}&0.18 & 0.02&10.99&$<2\times 10^{-16}$\\
\textbf{Intercept}& 0.21 & 0.02 &13.94 &$<2\times 10^{-16}$ \\
\hline
\hline
\end{tabular}
\end{adjustbox}
\label{tbl:linear_regression}
\end{center}
\end{table}

% \begin{table}
% \begin{center}
% \centering \caption {Linear Regression Model for how capability and intent contribute trust}
% \begin{adjustbox}{width=0.5\textwidth}
% \begin{tabular}{c c c c c c}
% \hline
% \hline
%  & \multicolumn{5}{c}{\textbf{Trust, Capability, Intent to Trust Decision}}\\
% \textbf{} & \textbf{coef.} & \textbf{std err} &  \textbf{z} & \textbf{$\boldsymbol{P>|z|}$} &\textbf{95\% CI}\\
% \hline
% \textbf{Capability}&0.5780  &0.016 & 35.981 &0.0001&[15.226,      18.780]\\
% \textbf{Intent}&0.1378 & 0.015 & 9.318 & 0.0001  &[0.109,       0.167]\\
% \textbf{Intercept}& 17.0030 & 0.906 &18.761& 0.0001 &[15.226,      18.780] \\
% \hline
% \hline
% \end{tabular}
% \end{adjustbox}
% \label{tbl:linear_regression}
% \end{center}
% \end{table}

% \begin{figure}
% \centerline{\includegraphics[scale=0.4]{figure/h3-2-tci_dec}}
% \caption{Self-reported trust, capability and intent scores, grouped by trust decision.}
% \label{fig:tciTrustDecision}
% \end{figure}

\vspace{0.5em}
\noindent\textbf{H4: Both inferred capability and intention influence human decisions to trust the robot.}
Similar to H3 above, we conducted mixed-effects analysis to account for dependent measures. In this case, the dependent variable was the trust decision (i.e., whether to delegate the task to the robot). Our primary model included random intercepts for subjects, scenarios, and tasks. The model's main fixed effects were inferred robot capability and intention, as well as two additional variables: \emph{residuals} from the self-reported trust model in H3, and participant preference. The residuals represented potential factors/components of trust that are separate from capability and intent. Participant preference was a binary indicator variable that (depending on the task) captured whether the participant preferred the higher risk option (for Tasks 1 and 3), or preferred the higher accuracy option (for Task 2). 

Our results (summarized in Tbl. \ref{tbl:logistic_regression_parameters}) show that decision to trust was significantly affected by both capability ($b = 0.94, \textrm{SE}=0.32, p<10^{-2}$) and intention ($b = -1.80, \textrm{SE}=0.38, p<10^{-5}$). There was a significant interaction between intention and participant preference ($b = 3.99, \textrm{SE}=0.47, p<10^{-16}$), which can be interpreted as the importance of \emph{intention alignment}. For example, when subjects were risk-seeking, their decisions were positively associated with the degree of agreement that the robot was also risk-seeking. Conversely, when participants were risk-averse, their decisions were negatively associated with the robot's inferred risk-preference. 

The trust residuals were also significantly associated with the trust decision ($b = 2.46, \textrm{SE}=0.40, p<10^{-9}$), suggesting that other factors (e.g., robot appearance, environment or human related elements~\cite{Hancock2011}) played a role in the eventual decision. However, a model with only self-reported trust as the independent variable (and otherwise the same as above), resulted in larger AIC and BIC scores (See Tbl.~\ref{tbl:aicbic} where $\triangle$AIC and $\triangle$BIC are differences from the best model). This suggests that a multidimensional mental construct is applied to trust-based decisions. Indeed, removing different components from the initial decision model resulted in poorer quality candidate models, as shown in Tbl. \ref{tbl:aicbic}.

\begin{table}
\begin{center}
\centering \caption {Mixed Effects Logistic Regression Model for Trust Decisions.}
\begin{adjustbox}{width=0.48\textwidth}
\begin{tabular}{l c c c c}
\hline
\textbf{} & \textbf{Coef.} & \textbf{SE} & \textbf{z value} &\textbf{Pr$(>|z|)$}\\
\hline
\textbf{Capability}& 0.94 &0.32&2.93&$3.39 \times 10^{-3}$\\
\textbf{Intention}&-1.80&0.38&-4.69& $2.68\times 10^{-6}$\\
\textbf{Trust Residual} & 2.46 & 0.40&6.20& $5.83\times 10^{-10}$ \\
\textbf{Preference}&-2.35&0.27&-8.86& $<$  $2\times 10^{-16}$\\
\textbf{Preference:Intent}&3.99&0.47&8.55&$<$ $2\times 10^{-16}$\\
\textbf{Intercept}& 1.36&0.32&4.23&2.39$\times 10^{-5}$ \\
\hline
\hline
\end{tabular}
\end{adjustbox}
\label{tbl:logistic_regression_parameters}
\end{center}
\end{table}

\begin{table}
\caption{$\triangle$AIC and $\triangle$BIC scores for four trust decision models. The ``Complete Model'' is the full trust decision model (described in the text) and achieves the best (lowest) AIC/BIC scores, compared to the alternative models. 
}
\begin{center}
\begin{adjustbox}{width=0.40\textwidth}
\begin{tabular}{l c c}
\hline
\textbf{Model} & \textbf{\textit{$\triangle$AIC }}&\textbf{\textit{$\triangle$BIC }}  \\
\hline
\hline 
\textbf{Complete Model}& 0&0\\
\textbf{\quad Without Intention}& 83.58& 66.77\\
\textbf{\quad Without Capability}& 39.29 &28.09\\
\textbf{\quad Without Trust Residuals}& 37.41 &31.80
\\
\textbf{Trust Only}& 81.98 & 59.58\\
\hline
\hline 
\end{tabular}
\end{adjustbox}
\label{tbl:aicbic}
\end{center}
\end{table}

\begin{figure}
\centerline{\includegraphics[width=0.45\textwidth]{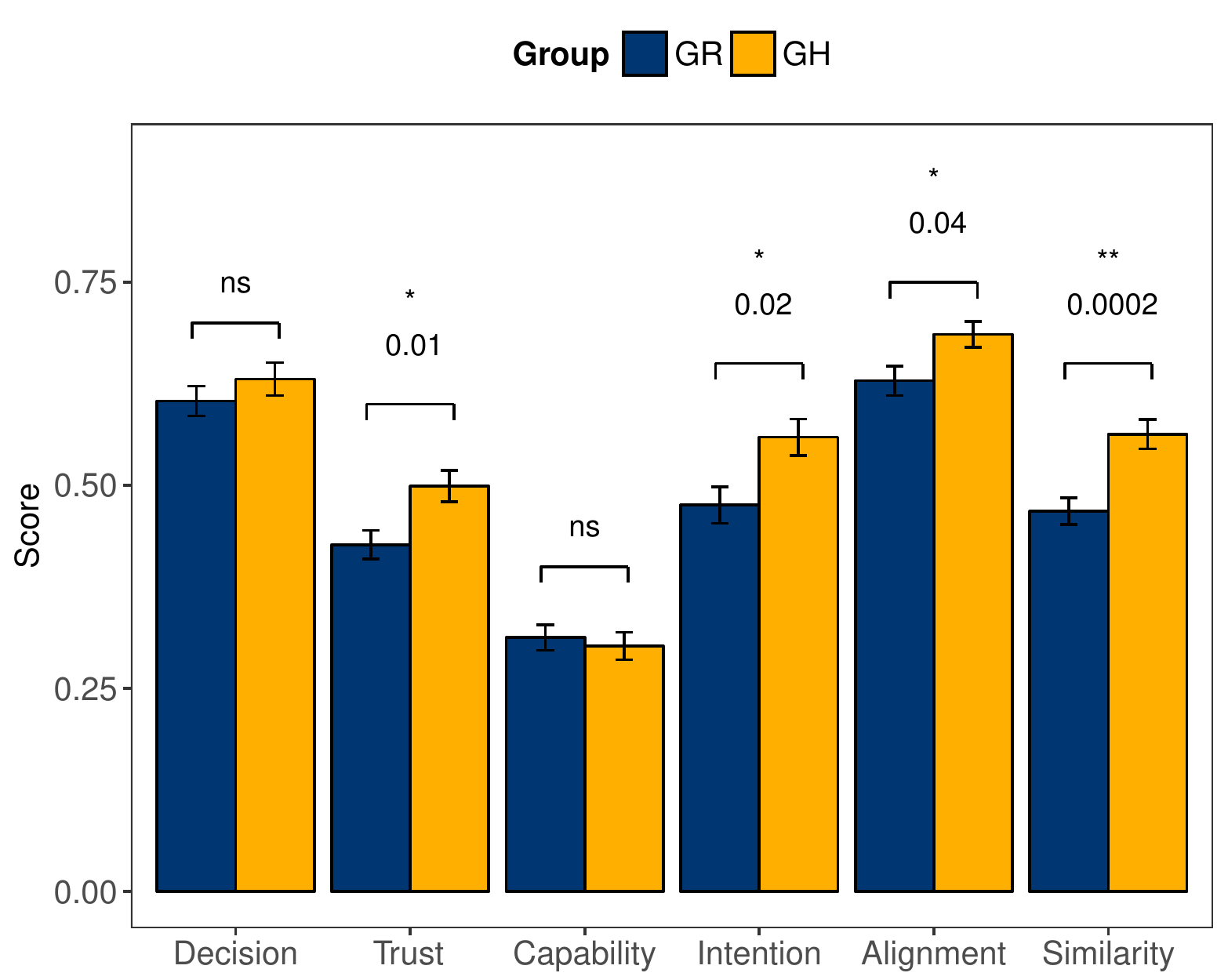}}
\caption{Difference in dependent variables between the simulated human group (GH) and simulated robot group (GR), with $p$-values shown above the bars. Participants reported significantly higher trust, intention, alignment and similarity in GH compared to GR.}
\label{fig:GHGRcomparisons}
\end{figure}

\vspace{0.5em}
\noindent\textbf{H5: People are more trusting of the simulated human agents, rather than the simulated robot agents.}
Fig. \ref{fig:GHGRcomparisons} shows the participants paired with simulated human agent (GH) reported significantly higher trust ($t(398) = -2.77 , p =0.01 $), inferred intent scores ($t(398) = -2.63 , p =0.02 $), inferred alignment ($t(398) = -2.36 , p =0.04 $), and similarity in decision-making ($t(398) = -3.91 , p < 10^{-2}$). However, differences in reported capability were not significant; potentially, participants considered capability as a property of the UAV since the other human player only controlled the location choice. Interestingly, differences in eventual trust decisions were also not statistically significant ($t(393.67) = -0.99, p=0.32$); there is marginal evidence that participants tended to delegate more to the HC-HR robot ($t(98) = -1.57, p=0.12$), but not to the other robot types ($p>0.5$). In other words, humans tend to report higher trust in other people (rather than robots) and believe people share their decision-making objectives. But at the same time, the decision to trust appears to remain dependent on observed behavior. These mixed results suggest more investigation is needed to further elucidate the differences between inter-human and human-robot trust.

%hese results support previous findings that people tend to collaborate more with human agents rather than robot agents, even when both act similarly~\cite{Collins2016,hripd}. Moreover, participants in GH predicted the agent to have more aligned ($M_{GR}=0.62,M_{GH}=0.68;U=17572.5,p=0.017)$), with a higher similarity in decision making ($M_{GR}=0.47,M_{GH}=0.56;U=15762.5, p=0.001$). 

% Fig. \ref{fig:GHGRcomparisons} shows the participants paired with simulated human agent (GH) were more likely to decide to delegate ($M_{GR}= 0.60, M_{GH} = 0.63; U = 2802000.0 p=0.02)$) and reported significantly higher trust and intent scores ($M_{Trust-GR}=0.42, M_{Trust-GH} =0.49 ; U =17146.5, p=0.006$ and $M_{Intent-RH} =0.47, M_{Intent_GH}=0.56; U= 16982.5, p = 0.004$). These results support previous findings that people tend to collaborate more with human agents rather than robot agents, even when both act similarly~\cite{Collins2016,hripd}. Moreover, participants in GH predicted the agent to have more aligned ($M_{GR}=0.62,M_{GH}=0.68;U=2715600.0,p<0.001)$), with a higher similarity in decision making ($M_{GR}=0.47,M_{GH}=0.56;U=15762.5, p=0.001$). These results suggest humans tend to believe that other humans share their decision-making processes, but this was not extended to robots.

\section{Discussion}
The results in the previous section show that both inferred capability and intent influence human decisions to trust the robot. Humans appear to require that a robot demonstrate that it has similar intent (e.g., risk or accuracy preference in our setup), in addition to having the capability to execute the task successfully. However, inferred robot intention and capability are by themselves insufficient; other factors---captured by the overall trust---contribute towards trust-based decision-making. These findings suggest that humans make trust-based decisions using \emph{rich mental representations} of robots, rather than relying solely on overall trust. %capability or intent. 

Trust in the simulated human agents and simulated robot agents were qualitatively similar, albeit with clear quantitative differences. Humans reported higher trust and inferred intent-alignment scores when partnered with simulated humans, which echo prior findings~\cite{Collins2016, Madhavan2007}. This points to a potential difference in people's prior expectations when working with humans versus machines.
%Since each agents type had the same behavior regardless of the group, this difference was not caused by observations of performance. %Rather, we hypothesize humans employ prior notions when collaborating with humans, which are not transferred to robots. 
%Furthermore, our work shows that human trust in robots is complex; it is dissimilar to trust in simple machines/automation (which is largely based on capability), but also not entirely the same as trust in humans, as shown by H5.
However, our failure to find statistically significant differences in decisions to trust suggests that trust decisions might vary for different agents, e.g., according to perceived similarity to human beings or specialization, and observed behavior. %Further study of prior expectations of various robot types for the purpose of designing proper human-robot interaction.

Taken together, these results have implications for the development of computational trust models~\cite{optimo,SohRSS18} and robots that consider human trust in their decision-making process (e.g.,~\cite{chen}). In particular, if a robot aims to \emph{predict} human behavior and act accordingly, tracking overall trust is inadequate when working in multiple task contexts. Instead, humans appear to internally represent agent capability and intention, allowing them to generalize appropriately to new scenarios. %, and has important implications towards designing human-centric robots. %For example, robots can use estimates of a user's trust as a factor in their decision-making process~\cite{chen}, or as a \textit{goal} (e.g., ``I should act in a manner to calibrate my user's trust in my capabilities and intentions"). 
Our findings also advise the practical aspects of human-centric robots. For example, calibrating user inferred capability and intention in assistive robots (such as smart wheelchairs~\cite{Soh2013,soh2015learning}) may encourage adoption and proper usage.

\section{Conclusions}
% notes for Yaqi and Indu: the first sentence of a strong conclusion should be WHY this work is important. Then go on to what we did.
Trust in autonomous robots will become increasingly important as advancements in robotics are putting robots on our streets, in our factories, and in our homes. It is thus critical to study and model how people trust and delegate tasks to robots.
In this paper, we presented human subjects with simulated task-delegation choices, and found that human decisions to delegate control in novel task contexts depend not only on overall trust in the robot collaborator, but also on estimations of robot capability and intention. 
That is, human trust in robots qualitatively mirrors human trust in other humans and is \textit{multifaceted}, consisting of at least two important facets: capability and intention.  

% Desmond: This following paragraph should go into the discussion!!! Not the conclusion! I removed the "limitations" bit, but rephrased the future work bit into the paragraph below.
%One of the limitations of the current work is that he number of tasks in our experiment is limited, and therefore is hard to model quantitatively. As for the next step, more tasks could be studied to establish the connection between task features to capability and intent transfer. \par

Our results add to a rich literature on factors that influence trust in robots (e.g.,~\cite{Hancock2011,Wang2016a}) and subsequent decision-making~\cite{chen,Robinette2017,Huangb}. Future work should examine other potential facets of human-robot trust, and provide empirical evidence for transfer across a wider range of tasks contexts. We envision a more complete theory of human-robot trust would contribute towards a collaborative trust-based society comprising both human and robot agents.  %bright (and trustworthy) future. % can also make the last line *even* more snazzy. 

%\textbf{\color{red} We need to cut the length down to 8 pages of text}

%Therefore we argue that the algorithms includes human mental models of robots should model both facets, capability and intention.\par

%\section*{Acknowledgements}
%This work was partially funded by a MoE Tier 1 

%\textbf{Future work: when the decision is hidden}

\balance
\bibliography{HRI}
\bibliographystyle{IEEEtran}

%\section*{Acknowledgment}

\end{document}